# Zero-shot self-supervised learning of single breath-hold magnetic resonance cholangiopancreatography (MRCP) reconstruction

Jinho Kim[1,2], Marcel Dominik Nickel[2], Florian Knoll[1]

[1]Department Artificial Intelligence in Biomedical Engineering, Friedrich-Alexander-Universität Erlangen-Nürnberg, Erlangen, Germany

[2]Research and Clinical Translation, Magnetic Resonance, Siemens Healthineers AG, Erlangen, Germany

# Abstract

**Purpose**:

To investigate the feasibility of zero-shot self-supervised learning reconstruction for reducing breath-hold times in magnetic resonance cholangiopancreatography (MRCP).

**Methods**:

Breath-hold MRCP was acquired from 11 healthy volunteers on 3T scanners using an incoherent k-space sampling pattern, leading to 14-second acquisition time and an acceleration factor of R=25. Zero-shot reconstruction was compared with parallel imaging of respiratory-triggered MRCP (338s, R=3) and compressed sensing reconstruction. For two volunteers, breath-hold scans (40s, R=6) were additionally acquired and retrospectively undersampled to R=25 to compute peak signal-to-noise ratio (PSNR).

To address long zero-shot training time, the $n + m$ full stages of the zero-shot learning were divided into two parts to reduce backpropagation depth during training: 1) $n$ frozen stages initialized with $n$-stage pretrained network and 2) $m$ trainable stages initialized either randomly or $m$-stage pretrained network. Efficiency of our approach was assessed by varying initialization strategies and the number of trainable stages using the retrospectively undersampled data.

**Results**:

Zero-shot reconstruction significantly improved visual image quality over compressed sensing, particularly in SNR and ductal delineation, and achieved image quality comparable to that of successful respiratory-triggered acquisitions with regular breathing patterns.

Improved initializations enhanced PSNR and reduced reconstruction time. Adjusting frozen/trainable configurations demonstrated that PSNR decreased only slightly from 38.25 dB (0/13) to 37.67 dB (12/1), while training time decreased up to 6.7-fold.

**Conclusion**:

Zero-shot learning delivers high-fidelity MRCP reconstructions with reduced breath-hold times, and the proposed partially trainable approach offers a practical solution for translation into time-constrained clinical workflows.

# Introduction

Magnetic resonance cholangiopancreatography (MRCP) is a non-invasive imaging technique used to visualize the biliary and pancreatic ductal systems, playing a critical role in the diagnosis of hepatobiliary diseases[1–4]. Traditionally performed using 2D thick-slab acquisitions, MRCP has evolved toward high-resolution 3D acquisitions to provide comprehensive anatomical detail[5]. However, this transition has come at the cost of significantly longer acquisition times, which increase susceptibility to motion artifacts and reduce image quality, particularly in free-breathing with uncooperative patients[6–8].

To mitigate these effects, respiratory-triggered acquisitions in free-breathing have been widely adopted. Techniques like prospective acquisition correction (PACE)[9] allow for improved motion suppression by synchronizing data acquisition with the patient's respiratory cycle, offering clearer anatomical contours and better patient tolerance. However, respiratory-triggered MRCP still suffers from prolonged and unpredictable scan durations, especially in patients with irregular or shallow breathing patterns[8,10–12], resulting in blurry images.

Breath-hold MRCP offers an alternative approach by acquiring data during a short breath-hold period, thereby eliminating respiratory motion artifacts[13–20]. $\ell_1$-wavelet compressed sensing reconstruction[21] was proposed for breath-hold MRCP due to the intrinsic sparsity of biliary structures[17–20]. However, it has been reported that it often fails to depict ductal details[19], which limits diagnostic confidence in assessing strictures or dilations. It also requires breath-hold durations of approximately 20s[10,14,15], which may not be tolerable for many patients, especially pediatric, elderly, or critically ill individuals[22]. To address these limitations, various strategies have been proposed, such as modifying the compressed sensing protocol by reducing the field-of-view[19] and training patients in breath-holding techniques before scanning[20,22,23]. However, reducing the field-of-view may fail to adequately cover the region of interest, and training patients before every scan is not trivial. Therefore, optimizing the k-space sampling patterns[24,25], such as a combination of equidistance and incoherent random undersampling offers a more practical and effective solution.

Deep learning-based MRI reconstructions have shown convincing results in accelerated MRI, particularly through physics-driven unrolled networks that integrate data fidelity and learned regularization[26–29]. These models typically require a large number of fully sampled training examples[26,27], which are difficult to obtain for applications in the abdomen like MRCP, where longer acquisitions are increasingly likely to be corrupted by motion artifacts. Yaman et al. proposed self-supervised deep learning reconstruction methods to eliminate the need for a fully sampled ground truth (self-supervised learning via data undersampled, SSDU[28]) and even to remove the need for training data at all, by training a model for a single specific acquisition (zero-shot self-supervised learning[29]).

Several scan-specific reconstruction approaches, that learn directly from the acquired data without large external databases, have been proposed. The zero-shot self-supervised approach shares the data-specific nature of these methods but differs fundamentally by combining physics-guided unrolled optimization with self-supervised loss partitioning. Unlike RAKI[30] or Residual-RAKI[31], it operates in the image domain with explicit data consistency terms rather than direct k-space interpolation. In contrast to Deep Image Prior[32] or implicit neural representations[33], it does not rely on the implicit bias of network architecture alone but enforces data consistency with measured data through the MRI encoding operator. Thus, our work bridges model-based deep learning and scan-specific adaptation, demonstrating zero-shot self-supervised reconstruction for highly accelerated 3D breath-hold MRCP.

In this work, we reduced the breath-hold duration of MRCP acquisitions and applied zero-shot learned reconstruction to improve image quality over conventional compressed sensing without relying on training datasets. One critical downside of zero-shot learning is the long training time of several hours per scan [29,34], which makes it infeasible for use in clinical practice. We developed a training strategy that leverages a pretrained reconstruction backbone by freezing the early stages of a self-supervised network and updating only the last trainable stages during zero-shot training. This reduces backpropagation depth and computation time with only a minimal trade-off in image quality.

# Methods

## Zero-shot self-supervised learning for MRCP reconstruction

Zero-shot reconstruction is designed for subject-specific self-supervised learning using only a single data[29]. In zero-shot learning, the acquired sampling pattern $\Omega$ is subdivided into three subsets of $\mathrm{T}$ for training, $\Lambda$ for loss, and $\Gamma$ for self-validation, and the available measurement samples from a single dataset are partitioned as:

$$\Omega = \mathrm{T} \sqcup \Lambda \sqcup \Gamma, \tag{1}$$

where $\sqcup$ denotes a disjoint union, i.e., $\mathrm{T}, \Lambda$ and $\Gamma$ are mutually exclusive to each other.

Each volumetric acquisition can be decoupled in readout direction using a Fourier transformation. The resulting $D$ readout positions, indexed by $j$, correspond to 2D k-space datasets in the phase encoding plane. This approach allows for more training samples and reduces memory requirements, assuming a neural network-based regularization that correlates pixels only in the phase encode directions.

For each readout index $j \in \{1, \ldots, D\}$, multiple disjoint mask pairs $\left(\mathrm{T}_k^j, \Lambda_\mathrm{k}^j\right)$ are generated within $\Omega^\mathrm{j} \setminus \Gamma^\mathrm{j}$, where $k \in \{1, \ldots, K\}$. These define the training dataset for zero-shot learning as:

$$\left(y_\mathrm{T}^{jk}, y_\Lambda^{jk}, y_\Gamma^\mathrm{j}\right) : j \in \{1, \ldots, D\}, k \in \{1, \ldots, K\}, \tag{2}$$

where $y$ is multi-coil undersampled k-space data. Zero-shot reconstruction then minimizes the loss function

$$\underset{\theta,\lambda}{\operatorname{argmin}} \frac{1}{DK} \sum_{j=1}^{D} \sum_{k=1}^{K} \mathcal{L}\left(y_\Lambda^{jk}, A_\Lambda^{jk}\left(f\left(y_\mathrm{T}^{jk}, A_\mathrm{T}^{jk}; \theta, \lambda\right)\right)\right), \tag{3}$$

where $f(\cdot)$ is the output reconstruction image of the unrolled network parametrized with $\theta$, and $A$ the encoding operator that contains undersampling, Fourier transform, and coil sensitivities. The regularization parameter $\lambda$ is learned during training, and $\mathcal{L}(\cdot)$ denotes a $\ell_1 - \ell_2$ loss function[28].

The long reconstruction times of zero-shot learning arise because deep neural network models must be trained individually for each subject, often requiring extensive optimization over many unrolled stages. Although transfer learning can accelerate convergence compared to random-initialized models[29], it still requires backpropagation through the entire unrolled architecture and repeated conjugate gradient-based data consistency computations at every stage (Figure 1a), which contributes heavily to memory and runtime costs.

To address this, we developed an alternative training approach that split a total $n + m$ stages of the unrolled network into $n$ frozen stages and $m$ trainable stages, where each stage was composed of a regularization block ($\mathcal{N}$) and a data consistency block (Figure 1a). The network $\mathcal{N}_\theta$ for the initial $n$ frozen stages were initialized with an $n$-stage pretrained network and their weights were frozen during zero-shot training. The weights for the $m$ trainable stages of the network $\mathcal{N}_{\theta'}$ were initialized either randomly or with an $m$-stage pretrained network and updated during training (Figure 1b). We used SSDU to obtain the pretrained backbone in this study, but in principle any training strategy can be applied. This design eliminates the need for backpropagation through the entire unrolled network. Since the frozen network deterministically maps the undersampled input ($x^0$) to an intermediate reconstruction ($x^n$), $x^n$ was cached during the first epoch to avoid redundant computation in the frozen part. Training then takes the form,

$$\underset{\theta,\lambda}{\operatorname{argmin}} \frac{1}{DK} \sum_{j=1}^{D} \sum_{k=1}^{K} \mathcal{L}\left(y_{\Lambda}^{jk}, A_{\Lambda}^{jk}\left(f'\left(y'^{jk}_{\mathrm{T}}, A_{\mathrm{T}}^{jk}; \theta', \lambda'\right)\right)\right), \tag{4}$$

where $y'$ denotes the precomputed input k-space of $\mathrm{x}^{\mathrm{n}}$ in Figure 1b, $f'$ indicates the model in the trainable stages, and $\lambda'$ is the trainable regularization parameter. In this setting, only the trainable stages are involved in gradient computation during training, significantly reducing backpropagation depth, memory consumption, and total training time.

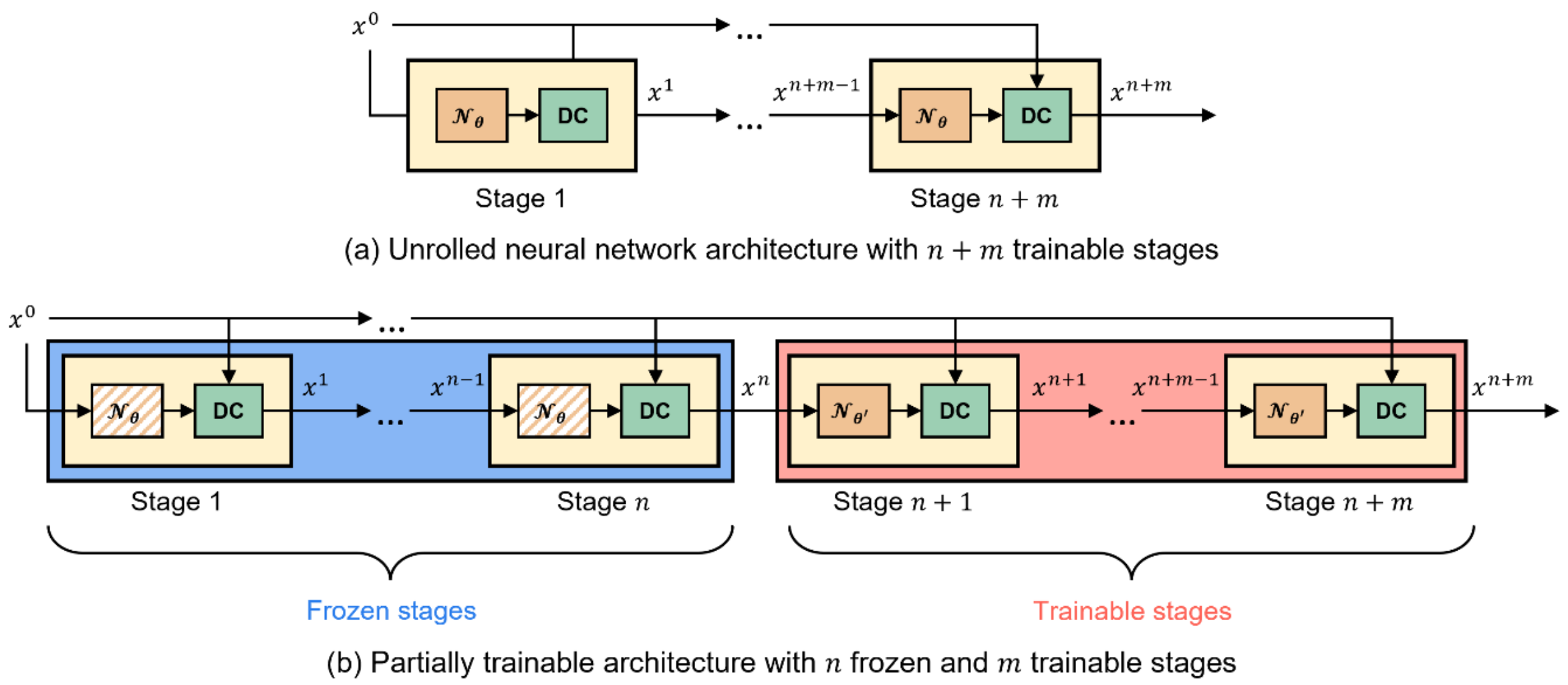

*Figure 1 (a) Unrolled neural network architecture with $n+m$ iterative stages. Each stage includes a regularization term ($\mathcal{N}$) and conjugate gradient-based data consistency (DC) block. (b) Partially trainable architecture using a pretrained reconstruction model: the initial $n$ stages are frozen (indicated by a patterned background), and later $m$ stages with a new neural network ($\mathcal{N}_{\theta'}$) are trainable.*

# Data

All participants in this study were informed about the study objectives and data handling procedures and subsequently provided written consent for participation and further data processing. There were two independent datasets used to train the pretrained model (SSDU) and the zero-shot learning, respectively.

## Dataset for pretrained model

The pretrained model was obtained using the subset of 3T acquisition from the raw dataset introduced in our previous study[27]. This dataset was collected from 31 healthy volunteers (19 males and 12 females) using 3T MRI scanners (MAGNETOM Vida and MAGNETOM Lumina, Siemens Healthineers AG, Forchheim, Germany) with a parallel imaging acceleration of R=2 and PACE triggering.

## Dataset for zero-shot learning

We acquired MRCP data from 11 healthy volunteers (nine males and two females) between June and November 2025 on 3T MRI scanners (MAGNETOM Lumina and MAGNETOM Prisma, Siemens Healthineers AG, Forchheim, Germany) using 12-channel body and 24-channel spine receive array coils. The age distribution of the volunteers ranged from 27 to 83 years, with a mean age of $43.5 \pm 19.8$ years. A 3D T2-weighted turbo spin-echo sequence (3D SPACE)[35] was used for both respiratory-triggered and breath-hold MRCP. Detailed protocol parameters are given in Table 1.

*Table 1 Protocol parameters for breath-hold MRCP and respiratory-triggered MRCP*

| | Breath-hold | Respiratory-triggered |
|---|---|---|
| Sequence | 3D T2-weighted TSE (3D SPACE) | |
| Acquisition plane | Coronal | |
| TR ($ms$) | $2000$ | $3165 - 6852^{*}$ |
| TE ($ms$) | $697$ | $703$ |
| TA ($s$) | 14 | $209 - 452^{*}$ |
| Acquired voxel size ($mm^3$) | $0.5 \times 0.5 \times 1.0$ | |
| Number of slices | $64$ | |
| Flip angles (°) | $100$ | $105 - 120^{*}$ |
| Number of signal averages | $1.0$ | $1.5$ |
| Triggering | N/A | PACE |
| Number of ACS lines | $24$ | |
| Total acceleration factor | $25$ | $3$ |

**Note**: The notation format for TR, TA, and Flip angles is Minimum-Maximum.
**Abbreviations**: TSE, turbo spin echo; TR, repetition time; TE, echo time; TA, acquisition time; N/A, not applicable; PACE, prospective acquisition correction; ACS, autocalibration signal.
**[*]** Variable depending on the volunteer.

Respiratory-triggered MRCP was acquired using a vendor-provided sequence with PACE triggering and parallel imaging acceleration with an undersampling factor of R=3 and 24 autocalibration lines. In this study, the goal of the respiratory-triggered acquisitions was to serve as an image quality reference against which we compared our approach. To obtain the best possible image quality, we obtained multiple acquisitions in the same volunteer until we were able to obtain an acquisition with a sufficiently regular breathing pattern. This approach resulted in two scans for volunteer #08, three scans for volunteers #07 and #10, and four scans for volunteer #11. A single scan was sufficient for remaining volunteers.

For breath-hold MRCP, we used a modified compressed sensing sequence[24,25], combining 2D Poisson-disk incoherent undersampling with equidistance undersampling and partial Fourier acquisition. Specifically, the 2D Poisson-disk sampling pattern included additional equidistant undersampling factors of R=3 and R=2 for phase- and partition-encoding directions, respectively. Partial Fourier acquisition was then applied, retaining 64% of k-space in the in-plane phase-encoding direction and 67% in the partition-encoding direction (Figure 2). This resulted in a total undersampling factor of R=25 and a single 14s breath-hold acquisitions. Reference scan was conducted separately with center 24 autocalibration lines on the phase encoding plane. For quantitative analysis, we additionally acquired 40-second breath-hold scans at R=6 from two of 11 volunteers in our study using the same sampling strategy and retrospectively undersampled to R=25.

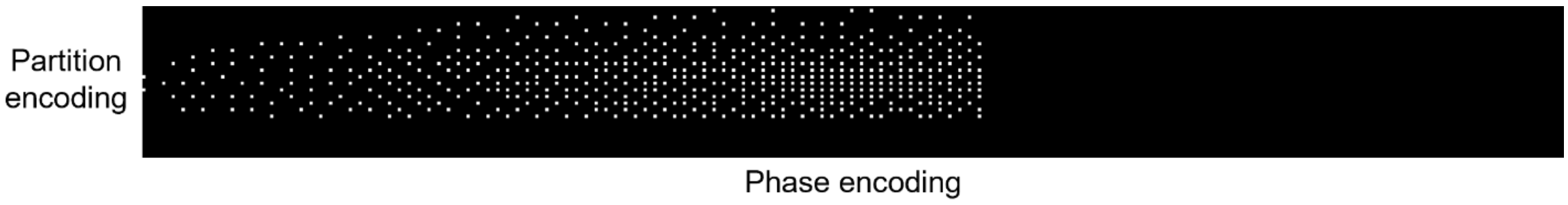

*Figure 2 Undersampling pattern for breath-hold MRCP, combining 2D Poisson-disk incoherent undersampling with partial Fourier undersampling, leading to a total acceleration factor of R=25 and 14s breath-hold scans. The x- and y-axis correspond to the in-plane phase encoding and partition encoding directions, respectively. The fully sampled readout (frequency encoding) direction is orthogonal to the phase-partition encoding plane (i.e., through-plane).*

# Data preparation

Raw data was extracted and converted into the ISMRMRD format[36] using the pyMapVBVD Python package[37]. We decoupled the fully sampled readout direction from the two phase-encoding directions by applying an inverse Fourier transform, followed by volume-wise normalization of the resulting stack of 2D k-space data using the maximum absolute value of all samples in the stack.

Coil sensitivity maps were pre-calculated using the ESPIRiT algorithm[38] using SigPy[39], with $24 \times 24$ fully-sampled central k-space data from the reference scan, a $5 \times 5$ kernel, and no background cropping in the image domain. The same coil sensitivity maps were consistently used across all model trainings and reconstructions to maintain consistency.

# Deep learning reconstructions

All training was conducted on a Linux system equipped with NVIDIA A100 40GB GPU devices. The source code for zero-shot learning reconstructions is available at https://github.com/JinhoKim46/ZS_BH-MRCP.

## Unrolled network

All deep learning reconstruction models in this study were based on unrolling an iterative algorithm. Each stage consisted of a learnable regularization module $\mathcal{N}$, and a conjugate gradient-based data consistency block (Figure 1a). $\mathcal{N}$ followed a ResNet[40] architecture, consisting of eight residual blocks with 64 channels, resulting in approximately 628 k trainable parameters. The weights of $\mathcal{N}$ were shared across all stages. All models were optimized using a normalized $\ell_1 - \ell_2$ loss in the k-space domain and trained with a cosine annealing learning rate schedule[41], starting from 0.0003. Notably, the learning rate was optimized to 0.0001 for volunteer #11 to ensure adequate image quality.

## Pretrained model

By acquiring multiple scans from some volunteers, the training dataset consisted of 39 data volumes for training, 4 volumes for validation, and 18 volumes for testing. 1D Gaussian sampling strategy was applied to split the acquired sampling set $\Omega$ into $\mathrm{T}$ and $\Lambda$ using a ratio $\rho_\Lambda = 0.4$[28]. We performed retrospective undersampling to R=25 using the sampling pattern shown in Figure 2. The number of unrolled stages of pretrained models was adjusted to match their target networks.

## Zero-shot learning

To train the zero-shot model, we followed the masking strategy of Yaman et al.[29], partitioning the acquired breath-hold MRCP k-space data Ω into three mutually exclusive subsets: Γ, Λ, and $\mathrm{T}$, with respective splitting ratios $\rho_\Gamma = 0.2$ and $\rho_\Lambda = 0.4$. The remaining data were assigned to $\mathrm{T}$. For each readout index $j$, $K = 10$ independent $(\mathrm{T}_k^j, \Lambda_k^j)$ pairs were generated from $\Omega^j \setminus \Gamma^j$. Models were using early stopping with a patience of three epochs and a minimum delta of $5 \times 10^{-3}$ based on the validation loss computed on $\Gamma$ to prevent overfitting during plateau phases. The regularization parameter $\lambda$ in Equation ( 3 ) was learned along with the model parameters $\theta$ to control the balance between the data consistency term and the learned prior within the unrolled network during training.

## Numerical experiments

We decided a total number of unrolled stages $n + m$ of 13 for our experiment. We compared zero-shot reconstructions of which trainable stages were initialized with either random weights or weights from an $m$-stage pretrained network to evaluate the effect of parameter initialization. We further investigated the effect of initialization for the frozen stages by defining the frozen part using either $\ell_1$-wavelet compressed sensing reconstruction or an $n$-stage pretrained network. In this setup, $x^n$ in Figure 1b represented the output produced by the selected initialization method. To quantify the quality-computation trade-off, we trained 13 configurations of zero-shot reconstructions for two retrospectively undersampled breath-hold MRCP. The configurations varied the number of $n$ frozen and $m$ trainable stages, resulting in 0/13 to 12/1 frozen/trainable stage combinations. The asterisk (*) was used as a label to indicate that the network for the trainable stages was initialized with the pretrained weights.

# Conventional reconstructions

For reference, we reconstructed our breath-hold MRCP acquisitions with $\ell_1$-wavelet compressed sensing using the SigPy[39]. The regularization parameter for compressed sensing was set to 0.008, as it provided visually optimal reconstructions by balancing data fidelity and regularization strength. To reconstruct the respiratory-triggered acquisitions, we used GRAPPA[42] using pygrappa[43].

# Evaluation

To evaluate reconstruction performance, we qualitatively compared our proposed methods against conventional compressed sensing, focusing on the visual fidelity of the reconstructed images. In the absence of fully sampled ground truth data, which is not feasible for clinical breath-hold MRCP acquisitions, we used our respiratory-triggered acquisitions as a surrogate reference. This choice was motivated by its widespread clinical use. We would like to note that the image quality of respiratory-triggered MRCP is strongly dependent on the regularity of the breathing pattern, and irregular breathing patterns are the main reasons why scans need to be repeated in practice. As described in the section on data acquisition, we repeated the triggered acquisition for some volunteers until we were able to achieve a sufficiently regular breathing pattern that could serve as a reference for image quality.

For the quantitative analysis, PSNR (dB) was computed from the retrospectively undersampled data at R=25 against its corresponding breath-hold acquisition at R=6 reconstructed with $\ell_1$-wavelet compressed sensing reconstruction.

The reconstruction time of zero-shot learning was measured over the entire reconstruction pipeline, including preprocessing, training and the final reconstruction.

# Results

Figure 3 shows a comparison of a successful respiratory-triggered acquisition (Figure 3a), an unsuccessful respiratory-triggered scan (Figure 3b), and our proposed 14s breath-hold acquisition (Figure 3c) for one of our volunteer scans (volunteer #10). The corresponding respiratory patterns are shown in Figures 3d and 3e. These results show that successful respiratory-triggering yields sharp and artifact-free images. Irregular or shallow breathing introduced motion blurring and navigator delays, which resulted in degraded duct visibility and prolonged the scan time from 360s (regular breathing) to 587s (irregular breathing). In comparison, the 14s breath-hold acquisition (Figure 3c) reduced motion artifacts in comparison to the failed triggering, but showed increased noise and reduced biliary visibility (e.g., pancreatic duct) in comparison to the successful triggering. Full respiratory traces of Figures 3d and 3e are available in Figure S1.

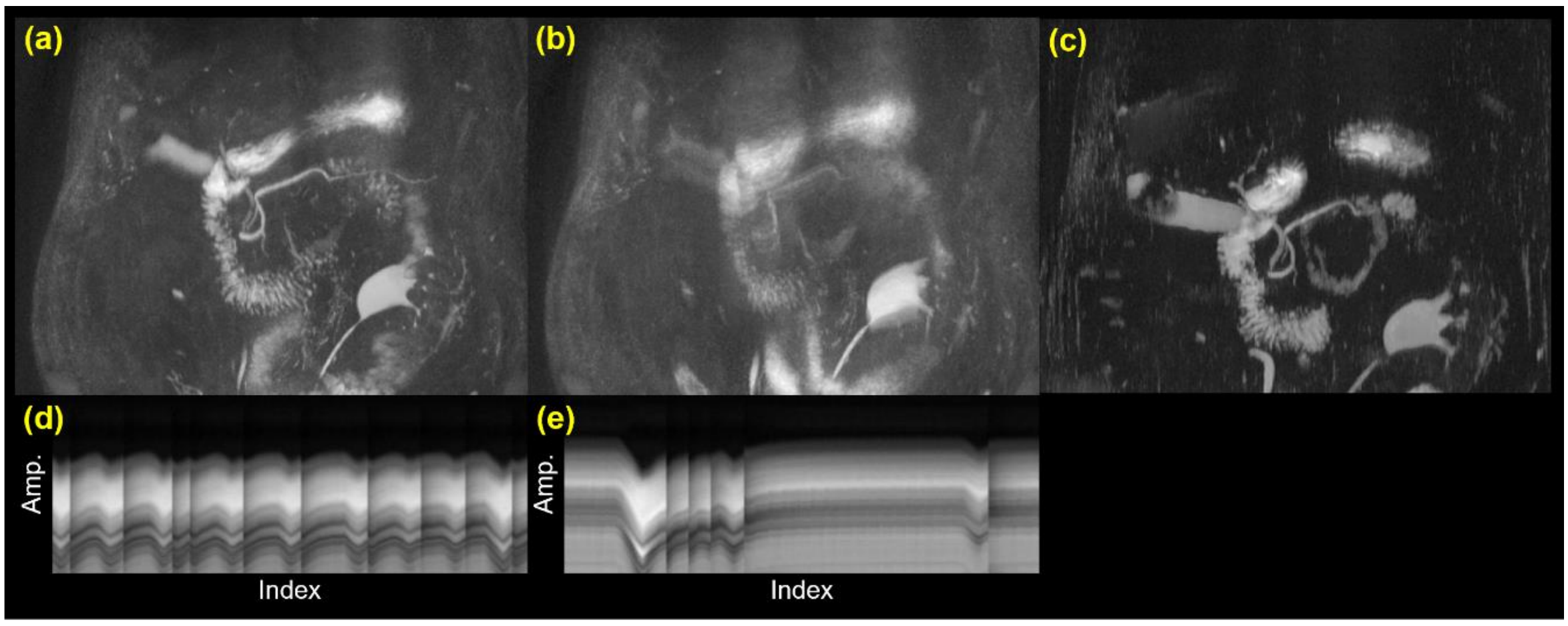

*Figure 3 MIP views showing the Impact of respiratory patterns on respiratory-triggered MRCP acquisition (a and b) and a corresponding breath-hold acquisition with zero-shot learning reconstruction of the same volunteer (c, volunteer #10). Subsets of the corresponding respiratory navigator signals are shown in panels d and e. The x-axis indicates signal sampling index of the acquired navigator, and the y-axis the amplitude of the respiratory signal.*

Figure 4 shows representative reconstructions with the PSNR and reconstruction times of zero-shot learning with different frozen/trainable stage configurations for two retrospectively undersampled acquisitions. The orange arrows highlight regions with notable differences in ductal visibility across reconstructions. The reference for PSNR in this study was the breath-hold MRCP at R=6 reconstructed with $\ell_1$-wavelet compressed sensing reconstruction. The asterisk (*) in the figure legend indicates that the network for the trainable stages is initialized with the pretrained weights. For zero-shot learning with random initialization across both datasets, the 0/13 configuration achieved PSNRs of 39.69 dB / 39.19 dB with reconstruction times of $102'12''$ / $223'21''$, whereas the 12/1 configuration yielded comparable PSNRs of 39.26 dB / 38.81 dB but reduced the reconstruction times to $24'19''$ / $24'20''$. For zero-shot learning with pretrained weight initialization, the 0/13* configuration produced PSNRs of 38.53 dB / 38.65 dB with reconstruction times of $81'35''$ / $121'23''$, while the 12/1* configuration achieved 38.63 dB / 37.30 dB with substantially shorter reconstruction times of $24'24''$ / $24'21''$.

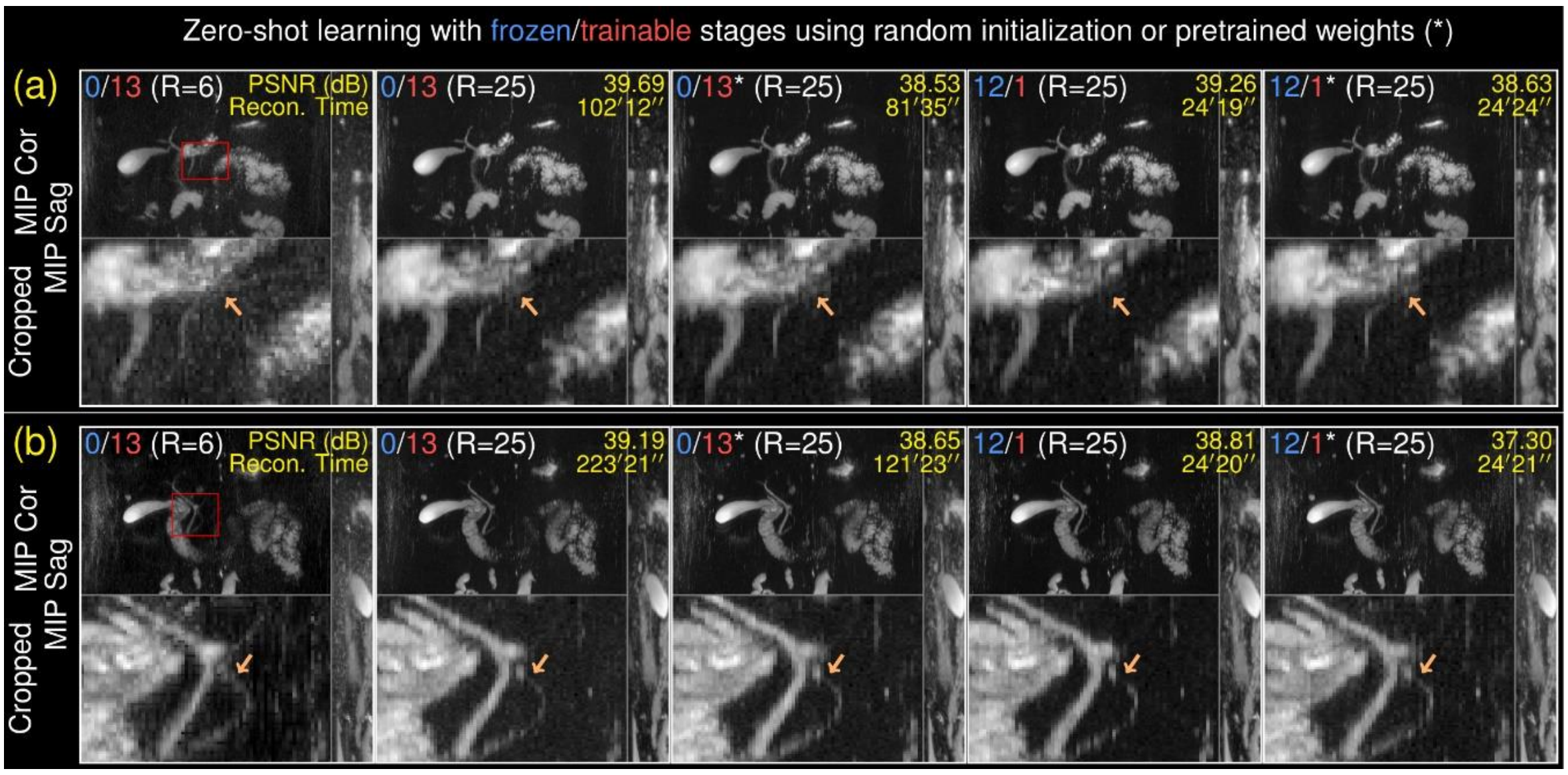

*Figure 4 Reconstructions results of retrospectively undersampled acquisition from (a) a 33-year-old male (volunteer #01) and (b) a 46-year-old male (volunteer #02) using zero-shot learning. Each block shows a breath-hold acquisition at R=6 and its retrospectively undersampling at R=25, reconstructed using different frozen/trainable configurations (0/13 and 12/1). The asterisk (*) means models where the trainable stages were initialized with pretrained weights. For each method, three visualizations are provided: a coronal maximum intensity projection (MIP, top left), a cropped coronal MIP focused on the region of interest (bottom left), and a sagittal MIP (right). The red box in the coronal MIP marks the ROI shown in the cropped view. Peak signal-to-noise ratio (PSNR) was computed against the original data at R=6. "Recon. Time" indicates the reconstruction time. The orange arrows highlight regions with notable differences in ductal visibility across reconstructions.*

Figure 5 depicts reconstruction results obtained using two different initialization strategies for the frozen stages. The frozen part of the network was defined either by an $\ell_1$-wavelet compressed sensing reconstruction (CS/1) or by an $n$-stage pretrained network (12/1), while the appended single trainable stage was initialized randomly. Two retrospectively undersampled acquisitions were reconstructed. The orange arrows indicate regions with visible differences in anatomical details, and the green arrows point to aliasing artifacts along the partition-encoding direction particularly for CS/1 in Figure 5b. Across both datasets, using the pretrained network for the frozen stages yielded higher PSNR values (39.26 dB / 38.81dB) compared with using the $\ell_1$-wavelet compressed sensing reconstruction (32.72 dB / 32.57 dB). The reconstruction times with 12 frozen stages ($24'19''$ / $24'20''$) were also shorter than those obtained with the compressed sensing initialization ($53'53''$ / $28'24''$).

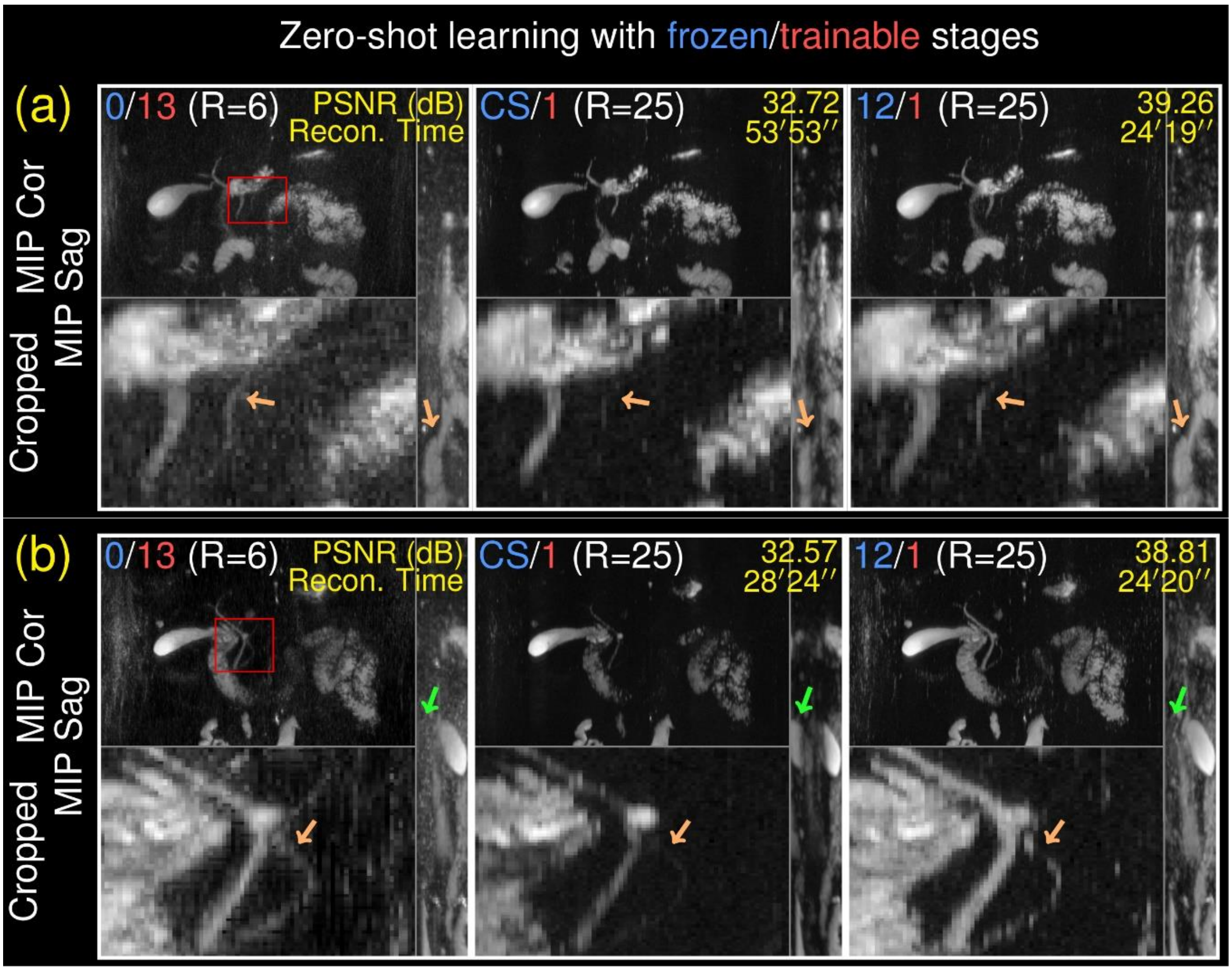

*Figure 5 Reconstructions results of different initializations for the frozen stages of zero-shot learning from (a) a 33-year-old male (volunteer #01) and (b) a 46-year-old male (volunteer #02). Each block shows a breath-hold acquisition at R=6 using the 0/13 configuration and its retrospectively undersampling at R=25, reconstructed using $\ell_1$-wavelet compressed sensing reconstruction (CS/1) and the 12-stage pretrained network (12/1) for the frozen parts with a single trainable stage configuration. For each method, three visualizations are provided: a coronal maximum intensity projection (MIP, top left), a cropped coronal MIP focused on the region of interest (bottom left), and a sagittal MIP (right). The red box in the coronal MIP marks the ROI shown in the cropped view. Peak signal-to-noise ratio (PSNR) was computed against the original data at R=6. "Recon. Time" indicates the reconstruction time. The orange arrows highlight regions with notable differences in anatomical details, and the green arrows point to the aliasing artifacts across reconstructions.*

Figure 6 plots average PSNR and reconstruction times from two retrospectively undersampled data across the various zero-shot configurations from 0/13* to 12/1*. The $m$-trainable stages were initialized with the $m$-stage pretrained networks, and this is marked with the asterisk (*) in the labels. The PSNR values remained across configuration shift (37.97 $\pm$ 0.54 dB, median: 38.25 dB) while the reconstruction times decreased approximately linearly.

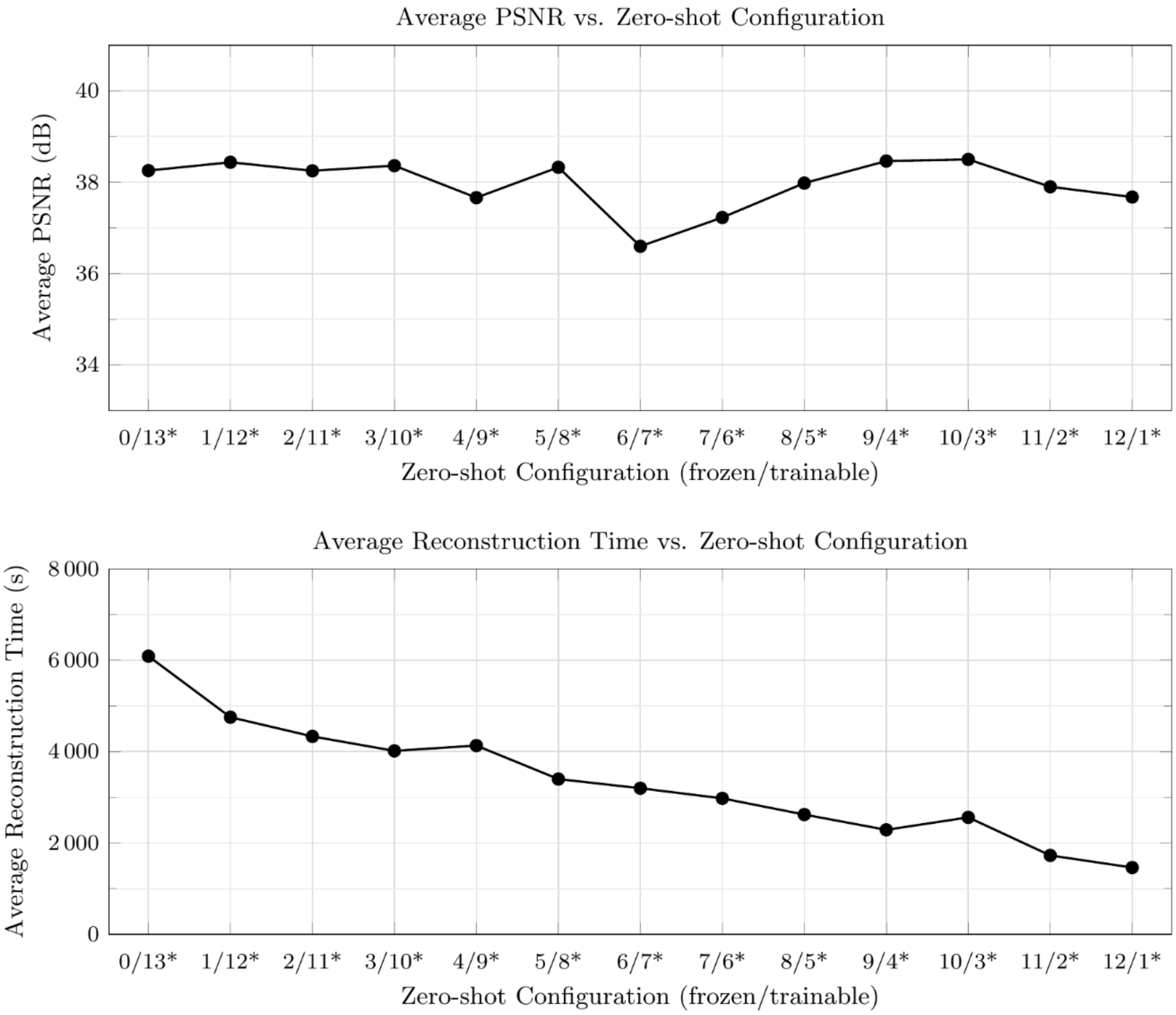

*Figure 6 Average peak signal-to-noise ratios (PSNR, top) and training times (bottom) of zero-shot models with varying frozen and trainable stage configurations across two retrospectively undersampled data. The asterisk (*) indicates that the $m$-trainable stages are initialized with the $m$-stage pretrained network. Each configuration represents the number of fixed and trainable stages in the zero-shot model. Results are averaged across two retrospectively undersampled datasets at R=25. PSNR was calculated against the original acquisition at R=6 using $\ell_1$-wavelet compressed sensing reconstruction.*

Figure 7 shows reconstruction results from a 28-year-old male (volunteer #07, Figure 7a) and a 64-year-old male (volunteer #08, Figure 7b). The figure compares the image quality of triggered acquisitions to 14s breath-hold acquisitions reconstructed with compressed sensing, a pretrained reconstruction model used for the frozen stages of ZS 12/1*, zero-shot learning with frozen/trainable configurations of 0/13* and 12/1*, respectively. Each reconstruction includes multiple visualizations (MIP Coronal, Cropped, and MIP Sagittal). The asterisk (*) indicates that the network for the trainable stages is initialized with the pretrained weights. With the exception of residual motion artifacts (highlighted by the blue arrow in Figure 7a), no breath-hold acquisition reaches the image quality of the (successful) triggered acquisition. However, ductal visibility (highlighted by orange arrows), contrast to noise ratio and absence of aliasing artifacts are superior for both zero-shot methods in comparison to compressed sensing and the pretrained reconstruction model. While 12/1* zero-shot training strategy shows a minor reduction in fine structural details relative to 0/13* zero-shot training, it still outperformed compressed sensing in terms of contrast to noise ratio and duct delineation. Full reconstruction results for additional subjects are provided in Figures S2 to S9.

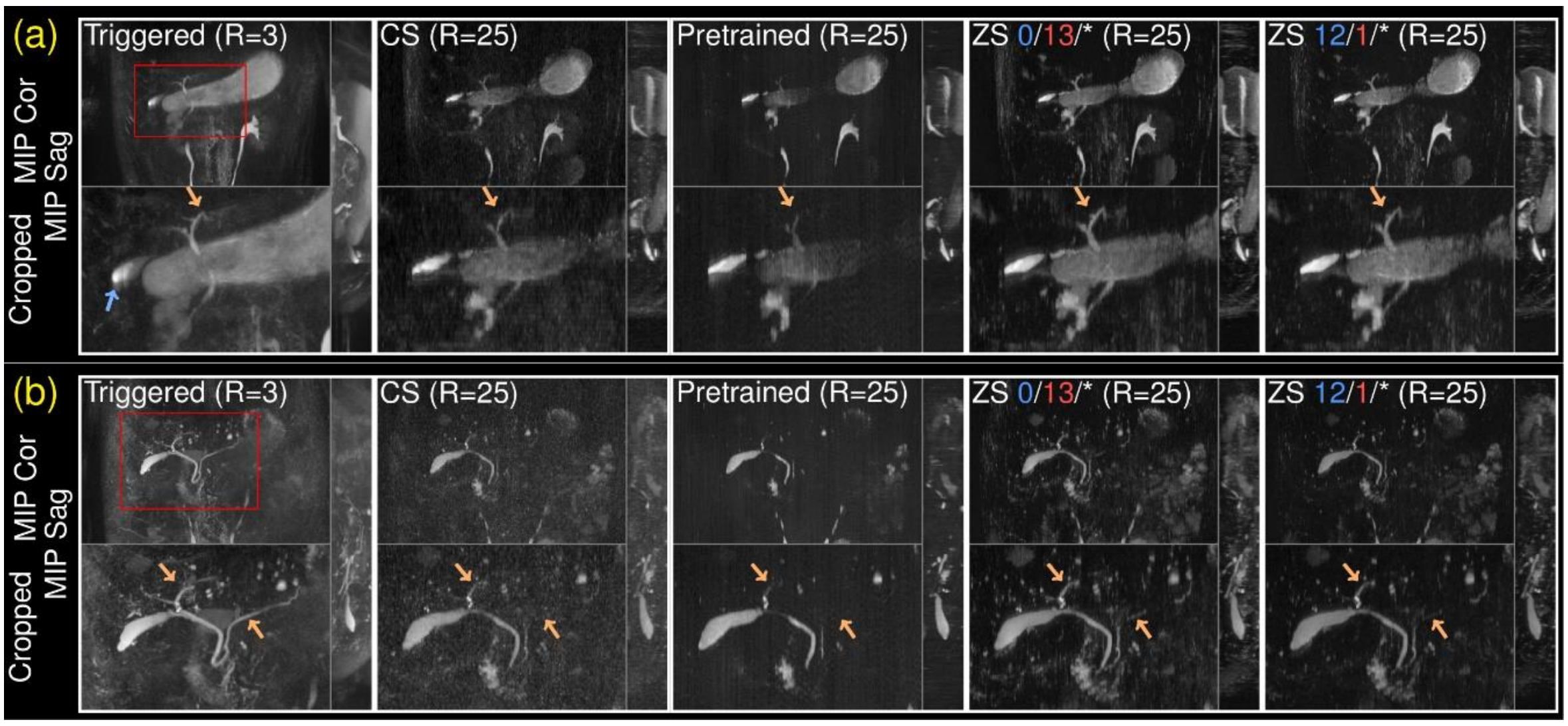

*Figure 7 Reconstruction results from (a) a 28-year-old male (volunteer #07) and (b) a 64-year-old male (volunteer #08). Each reconstruction block shows results from a triggered acquisition (Triggered) and a 14s breath-hold acquisition reconstructed with compressed sensing (CS), a pretrained reconstruction model used for the frozen stages of ZS 12/1* (Pretrained), zero-shot learning with frozen/trainable configurations of 0/13* and 12/1*, respectively. The asterisk (*) indicates that the network for the $m$-trainable stages is initialized with the $m$-stage pretrained weights. For each method, three visualizations are provided: a coronal maximum intensity projection (MIP, top left), a cropped coronal MIP focused on the region of interest (bottom left), and a sagittal MIP (right). The red box in the coronal MIP marks the ROI shown in the cropped view. The orange arrows highlight regions with notable differences in ductal visibility across reconstructions. The blue arrow indicates motion artifacts in the triggered acquisition.*

Our results demonstrated the occasional presence of residual aliasing artifacts in the slice encoding direction for the breath-hold acquisitions. Figure 8 (volunteer #01, 33-year-old male) shows a comparison of aliasing artifact suppression in the slice-encoding direction across techniques.

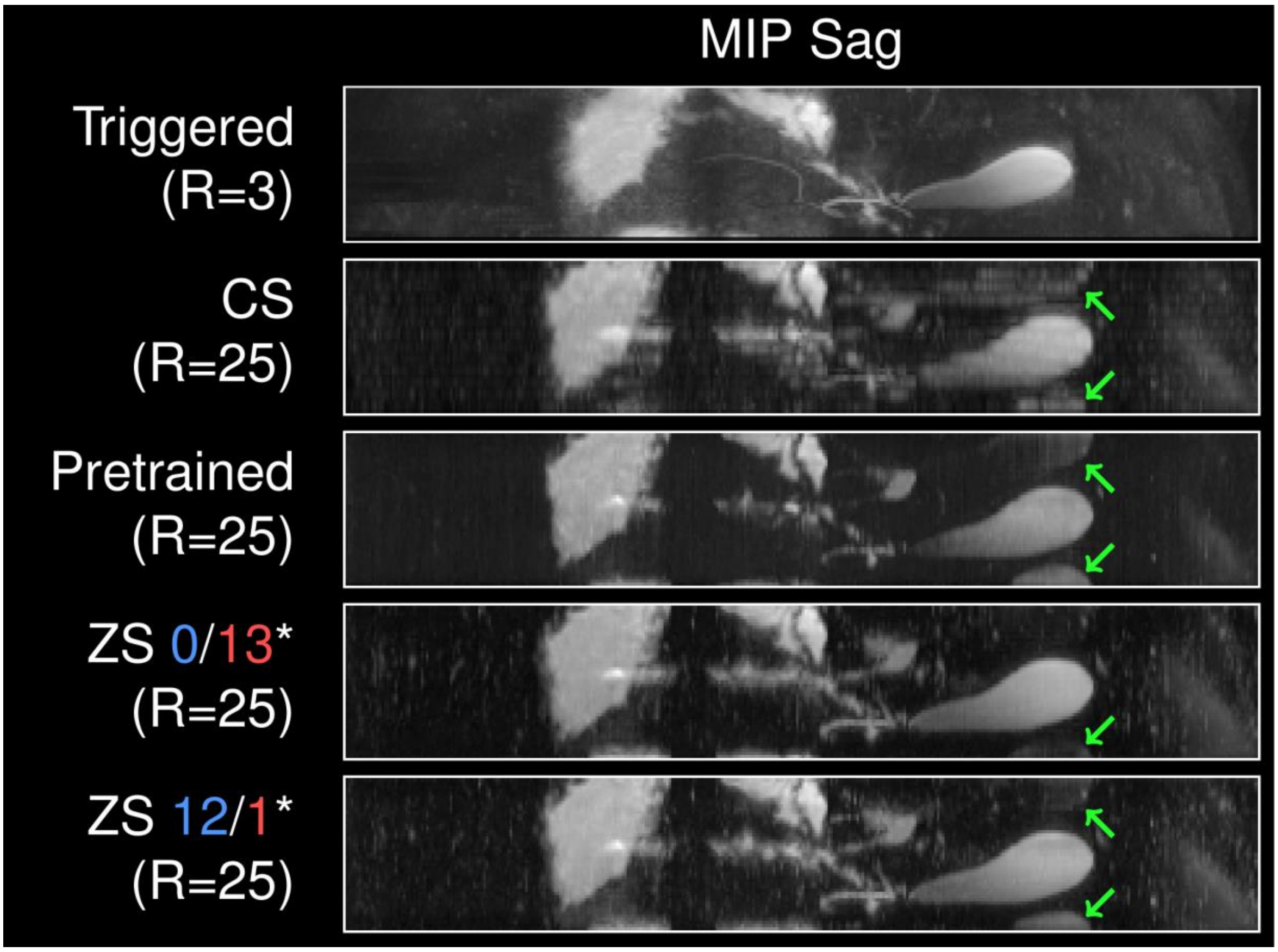

*Figure 8 Reconstruction results from a 33-year-old male (volunteer #01), demonstrating aliasing artifacts along the slice encoding direction in the sagittal MIP view. Each reconstruction block shows results from a triggered acquisition (Triggered) and a 14s breath-hold acquisition reconstructed with compressed sensing (CS), a pretrained reconstruction model used for the frozen stages of ZS 12/1* (Pretrained), zero-shot learning with frozen/trainable configurations of 0/13* and 12/1*, respectively The asterisk (*) indicates that the network for the trainable stages is initialized with the pretrained weights. Green arrows indicate areas where aliasing artifacts are observed.*

Table 2 shows reconstruction times of zero-shot training with different configurations of frozen/trainable stages for all 11 volunteers. The 0/13 configuration required $136'16''$ with random initialization and $78'46''$ with pretrained initialization, whereas the 12/1 configuration reduced the reconstruction times to $20'25''$ and $18'19''$, respectively. This corresponds to 6.7- and 4.3-fold speedup in reconstruction time for the 12/1 configuration relative to 0/13 for each random and pretrained initialization, respectively.

*Table 2 Reconstruction times of zero-shot learning with different configurations of frozen/trainable stages for all 11 volunteers in this study. The asterisk (*) indicates fine-tuned models where the $m$-trainable stages were initialized with $m$-stage pretrained weights.*

| Volunteer | Zero-shot configurations (frozen/trainable) | | | |
|---|---|---|---|---|
| | 0/13 | 0/13* | 12/1 | 12/1* |
| #01 | $122'14''$ | $77'39''$ | $19'21''$ | $19'33''$ |
| #02 | $131'31''$ | $74'00''$ | $22'08''$ | $15'32''$ |
| #03 | $190'34''$ | $77'47''$ | $19'22''$ | $19'15''$ |
| #04 | $203'47''$ | $112'10''$ | $17'53''$ | $17'31''$ |
| #05 | $115'29''$ | $77'03''$ | $19'05''$ | $19'18''$ |
| #06 | $107'48''$ | $75'51''$ | $18'02''$ | $17'54''$ |
| #07 | $129'07''$ | $75'26''$ | $17'24''$ | $17'36''$ |
| #08 | $121'58''$ | $69'22''$ | $19'04''$ | $19'35''$ |
| #09 | $121'28''$ | $76'31''$ | $19'45''$ | $19'27''$ |
| #10 | $83'47''$ | $74'15''$ | $24'57''$ | $16'48''$ |
| #11 | $171'22''$ | $76'28''$ | $27'37''$ | $19'00''$ |
| Average | $136'16''$ | $78'46''$ | $20'25''$ | $18'19''$ |

**Note**: The asterisk (*) indicates that the network for the $m$-trainable stages is initizliaed with the $m$-stage pretrained network.

# Discussion

The current state of the art in breath-hold MRCP requires breath-hold durations of around 20 seconds, which can be challenging for patients[44]. The goal of this study was to make breath-hold MRCP more feasible. By combining 2D Poisson-disc and Partial Fourier undersampling, we achieved an acquisition time of 14 seconds. This falls within the ideal range of 10–15 seconds and substantially improves the feasibility of breath-hold MRCP, even for populations with limited breath-hold capacity such as pediatric, elderly, or critically ill patients[23,44].

In the absence of a fully sampled ground truth, we used respiratory-triggered acquisitions as the reference for image quality in this study. While none of our breath-hold acquisitions reached the image quality of the triggered acquisition in the case of regular breathing patterns, we noticed that even young healthy volunteers reported discomfort from the prolonged and inconsistent acquisition times. In four volunteers, we had to repeat the respiratory-triggered acquisitions to achieve sufficient image quality.

For the quantitative evaluation, we undersampled 6-fold breath-hold acquisition retrospectively to the factor of R=25. Since the R=25 sampling pattern is a subset of the R=6 acquisition, we retrospectively undersampled the R=6 data by applying the R=25 sampling mask.

Our results demonstrated superior image quality of image reconstruction using zero-shot learning. Conventional compressed sensing resulted in poor depiction of ductal structures at such a high acceleration rate. Using a pretrained model trained with retrospective subsampling of triggered acquisitions also led to inferior image quality. Several systematic differences from our previous study[27] likely contributed to this performance discrepancy. These include a substantially higher acceleration factor (R=25 vs. R=6), the additional use of partial Fourier acquisition, and different sampling strategies (incoherent vs. equidistant).

Several types of artifacts were observed in our reconstruction results. First, since we decoupled the fully sampled readout direction and performed 2D reconstruction across the two phase-encoding directions, discontinuities between adjacent readout planes occasionally appeared. Second, the sampling pattern contained equidistance undersampling along the phase-encoding (R=3) and partition-encoding (R=2) directions. Given the relatively small number of partition-encoding lines, the acceleration factor of R=2 in this direction led to residual aliasing in the partition dimension. The 0/13 zero-shot reconstruction was able to suppress these aliasing artifacts effectively, whereas the 12/1 configuration showed limited ability to fully remove them. In contrast, conventional compressed sensing failed to suppress the aliasing artifacts and additionally exhibited wavelet-domain shrinkage artifacts characteristic of $\ell_1$-regularized reconstructions.

In our different initialization approaches for the trainable stages, pretrained initialization consistently converged faster than random. For the 0/13 configurations, the impact of initialization was minimal in both quantitative metrics and the visual appearance. In contrast, for the 12/1 configurations, pretrained-weight initialization resulted in better image quality, particularly in preserving the visibility of the pancreatic ducts (orange arrows in *Figure 4*). Furthermore, the experiment demonstrated that fine-tuning with the 12/1 configuration was more computationally efficient than fine-tuning with the 0/13 configuration, because it achieved comparable image quality to 0/13 zero-shot learning but significantly shorter reconstruction time.

Given that the network weights were shared across stages, the number of trainable parameters remained constant across all zero-shot configurations. Only the backpropagation depth varied with the number of trainable stages, leading to the approximately linear decrease in reconstruction time.

We would like to note that although partial Fourier acquisition was employed during data acquisition, no partial Fourier reconstruction techniques were applied. Future work may explore the integration of partial Fourier methods into deep learning reconstructions[45] to further enhance image quality. Moreover, subject-specific hyperparameter adjustment, such as learning rate tuning, may be necessary to optimize reconstruction performance across different cases. As an example, the learning rate for volunteer #11 was reduced to 0.0001 since the initial value of 0.0003 caused optimization instability and prevented convergence, whereas the default value was optimal to the other cases. With the reduced learning rate, zero-shot reconstruction for volunteer #11 achieved image quality and reconstruction time comparable to the other zero-shot models.

The main limitation for the clinical translation of zero-shot learning is the long computation time. This motivated the development of the partial trainable zero-shot training strategy, which reduces backpropagation depth by splitting a total number of stages into frozen and trainable stages. While this approach substantially decreases training time compared to conventional zero-shot learning, we acknowledge that the overall reconstruction time still remains longer than that of other non-data-specific deep learning–based methods. Various deep learning reconstruction models could be used in this framework. In this study, we selected SSDU[28] as the backbone due to its architectural similarity to zero-shot learning, and its self-supervised formulation, which makes it particularly well suited for MRCP where fully sampled data is unavailable[27].

Even though the participants of our study are from a representative age group (the oldest subject being 83 years old), one limitation is the exclusive use of healthy volunteer data. Therefore, our results do not allow to generalize our findings to pathologic cases. Also, the distribution of regular and irregular breathing patterns as well as consistency of breath-holds may vary in a clinical population with sick patients[20]. Furthermore, the image quality assessment was based solely on visual inspection by the authors, without any involvement of radiologists or a radiologist-led reader study. Future work should validate the clinical applicability of zero-shot methods using patient datasets and expert diagnostic assessments.

# Conclusion

The goal of our study was to shorten the breath-hold duration in MRCP, to provide an alternative to commonly used respiratory triggered acquisitions. Our results demonstrate that image reconstruction using zero-shot learning outperformed compressed sensing and the use of a pre-trained reconstruction model at the required acceleration factor to achieve a breath-hold duration of less than 15 seconds. These findings suggest that zero-shot learning is a promising reconstruction approach when large training datasets are unavailable.

We also introduce a zero-shot training strategy with reduced backpropagation depth that reduced the reconstruction time from $136'16''$ to $20'25''$ for random initialization and from $78'46$ to $18'19''$ for pretrained-weight initialization, with only a minor trade-off in reconstruction quality. We expect that this reduced training time can enable the integration of zero-shot learning into clinical workflows.

# Acknowledgements

Funding by the German Research Foundation (DFG) is gratefully acknowledged (projects 513220538, 512819079, 556079651; and project 500888779 of the RU5534 MR biosignatures at UHF). In addition, we gratefully acknowledge the scientific support and HPC resources provided by the Erlangen National High Performance Computing Center (NHR@FAU) of Friedrich-Alexander-University Erlangen-Nuremberg (FAU) under the NHR project b143dc. NHR funding is provided by federal and Bavarian state authorities. NHR@FAU hardware is partially funded by the German Research Foundation (DFG) – 440719683.

The authors thank Hyeeun Park, Majd Helo, Marc Vornehm, and Yannik Ott for their contributions to dataset collection.

# Data availability statement

A sample data that supports the findings of this study are openly available at https://doi.org/10.5281/zenodo.16731625.